\documentclass[onefignum,onetabnum]{siamart190516}

\usepackage{lipsum}
\usepackage{amsfonts}
\usepackage{graphicx}
\usepackage{epstopdf}
\usepackage{algorithmic}
\ifpdf
  \DeclareGraphicsExtensions{.eps,.pdf,.png,.jpg}
\else
  \DeclareGraphicsExtensions{.eps}
\fi

\newsiamremark{remark}{Remark}
\newsiamremark{hypothesis}{Hypothesis}
\crefname{hypothesis}{Hypothesis}{Hypotheses}
\newsiamthm{claim}{Claim}

\headers{The Linear Algebra Mapping Problem}{C. Psarras, H. Barthels and P. Bientinesi}

\title{The Linear Algebra Mapping Problem\thanks{Submitted to the editors \chris{fill in}.
\funding{Financial support from the Deutsche Forschungsgemeinschaft (German Research Foundation) through grants GSC 111 and IRTG 2379 is gratefully acknowledged.}}}

\author{Christos Psarras\thanks{Aachen Institute for Advanced Study in Computational Engineering Science, RWTH Aachen University, Aachen, Germany
        (\email{psarras@aices.rwth-aachen.de, barthels@aices.rwth-aachen.de}).}
   \and Henrik Barthels\footnotemark[1]
   \and Paolo Bientinesi\thanks{Department of Computing Science, Ume{\aa} Universitet, Ume{\aa}, Sweden
    (\email{pauldj@cs.umu.se}).}}

\usepackage{amsopn}

\ifpdf
\hypersetup{
  pdftitle={The Linear Algebra Mapping Problem},
  pdfauthor={C. Psarras, H. Barthels and P. Bientinesi}
}
\fi

\definecolor{rwthblue}   {RGB}{  0,  84, 159}
\definecolor{rwthblue75}{RGB}{ 64, 127, 183}
\definecolor{rwthblue50}{RGB}{142, 186, 229}
\definecolor{rwthblue25}{RGB}{199, 221, 242}
\definecolor{rwthblue10}{RGB}{232, 241, 250}

\definecolor{rwthblack}  {RGB}{  0,   0,   0}
\definecolor{rwthblack75}{RGB}{100, 101, 103}
\definecolor{rwthblack50}{RGB}{156, 158, 159}
\definecolor{rwthblack25}{RGB}{207, 209, 210}
\definecolor{rwthblack10}{RGB}{236, 237, 237}

\definecolor{rwthdarkgray}{named}{rwthblack75}
\definecolor{rwthgray}{named}{rwthblack50}
\definecolor{rwthlightgray}{named}{rwthblack25}
\definecolor{rwthverylightgray}{named}{rwthblack10}

\definecolor{rwthmagenta}  {RGB}{227,   0, 102}
\definecolor{rwthmagenta75}{RGB}{233,  96, 136}
\definecolor{rwthmagenta50}{RGB}{241, 158, 177}
\definecolor{rwthmagenta25}{RGB}{249, 210, 218}
\definecolor{rwthmagenta10}{RGB}{253, 238, 240}

\definecolor{rwthyellow}  {RGB}{255, 237,   0}
\definecolor{rwthyellow75}{RGB}{255, 240,  85}
\definecolor{rwthyellow50}{RGB}{255, 245, 155}
\definecolor{rwthyellow25}{RGB}{255, 250, 209}
\definecolor{rwthyellow10}{RGB}{255, 253, 238}

\definecolor{rwthpetrol}   {RGB}{  0,  97, 101}
\definecolor{rwthpetrol75}{RGB}{ 45, 127, 131}
\definecolor{rwthpetrol50}{RGB}{125, 164, 167}
\definecolor{rwthpetrol25}{RGB}{191, 208, 209}
\definecolor{rwthpetrol10}{RGB}{230, 236, 236}

\definecolor{rwthturquoie}   {RGB}{  0, 152, 161}
\definecolor{rwthturquoise75}{RGB}{  0, 177, 183}
\definecolor{rwthturquoise50}{RGB}{137, 204, 207}
\definecolor{rwthturquoise25}{RGB}{202, 231, 231}
\definecolor{rwthturquoise10}{RGB}{235, 246, 246}

\definecolor{rwthgreen}  {RGB}{ 87, 171,  39}
\definecolor{rwthgreen75}{RGB}{141, 192,  96}
\definecolor{rwthgreen50}{RGB}{184, 214, 152}
\definecolor{rwthgreen25}{RGB}{221, 235, 206}
\definecolor{rwthgreen10}{RGB}{242, 247, 236}

\definecolor{rwthlightgreen}   {RGB}{189, 205,   0}
\definecolor{rwthlightgreen75}{RGB}{208, 217,  92}
\definecolor{rwthlightgreen50}{RGB}{224, 230, 154}
\definecolor{rwthlightgreen25}{RGB}{240, 243, 208}
\definecolor{rwthlightgreen10}{RGB}{249, 250, 237}

\definecolor{rwthorange}  {RGB}{246, 168,   0}
\definecolor{rwthorange75}{RGB}{250, 190,  80}
\definecolor{rwthorange50}{RGB}{253, 212, 143}
\definecolor{rwthorange25}{RGB}{254, 234, 201}
\definecolor{rwthorange10}{RGB}{255, 247, 234}

\definecolor{rwthred}  {RGB}{204,   7,  30}
\definecolor{rwthred75}{RGB}{216,  92,  65}
\definecolor{rwthred50}{RGB}{230, 150, 121}
\definecolor{rwthred25}{RGB}{243, 205, 187}
\definecolor{rwthred10}{RGB}{250, 235, 227}

\definecolor{rwthbordeaured}   {RGB}{161,  16,  53}
\definecolor{rwthbordeauxred75}{RGB}{182,  82,  86}
\definecolor{rwthbordeauxred50}{RGB}{205, 139, 135}
\definecolor{rwthbordeauxred25}{RGB}{229, 197, 192}
\definecolor{rwthbordeauxred10}{RGB}{245, 232, 229}

\definecolor{rwthviolet}  {RGB}{ 97,  33,  88}
\definecolor{rwthviolet75}{RGB}{131,  78, 117}
\definecolor{rwthviolet50}{RGB}{168, 133, 158}
\definecolor{rwthviolet25}{RGB}{210, 192, 205}
\definecolor{rwthviolet10}{RGB}{237, 229, 234}

\definecolor{rwthpurple}  {RGB}{122, 111, 172}
\definecolor{rwthpurple75}{RGB}{155, 145, 193}
\definecolor{rwthpurple50}{RGB}{188, 181, 215}
\definecolor{rwthpurple25}{RGB}{222, 218, 235}
\definecolor{rwthpurple10}{RGB}{242, 240, 247}

\newcommand{\chris}[1]{{\color{magenta} Chris: #1}}

\newcommand{\changed}[1]{#1}

\newcommand{\LT}{LT}

\newcommand{\DI}{DI}
\newcommand{\SYM}{SYM}
\newcommand{\SPD}{SPD}
\newcommand{\SPSD}{SPSD}

\usepackage{booktabs} 

\usepackage{amsmath,amsfonts}
\usepackage{amssymb}

\usepackage{xcolor,color,colortbl}
\usepackage{multirow}
\usepackage[most]{tcolorbox}
\tcbuselibrary{theorems}

\usepackage{tikz}
\usetikzlibrary{positioning}

\usepackage{pgfplots}
\usepackage{pgfplotstable}
\pgfplotsset{compat=newest}

\usepackage{footnote}
\usepackage{drawmatrix}
\usepackage[lofdepth,lotdepth]{subfig}
\usepackage{adjustbox}
\usepackage{fancybox}
\usepackage{bbding}
\usepackage{tabularx}
\usepackage{longtable}
\usepackage{float}

\newcommand{\matprod}{\;}

\newcommand{\dims}[2]{$\mathbb{R}^{#1 \times #2}$}
\newcommand{\dimms}[3]{$#1 \in \mathbb{R}^{#2 \times #3}$}
\newcommand{\order}[1]{$\mathcal{O}(#1)$}
\newcommand{\yes}{\checkmark}
\newcommand{\no}{\footnotesize{$-$}}
\newcommand{\na}{n.a.}
\newcommand{\pluseq}{\mathrel{+}=}
\newcommand{\symone}{$\dagger$}
\newcommand{\symtwo}{$\ddagger$}
\newcommand{\exdims}[3]{$#1 \in \mathbb{R}^{#2 \times #3}$}

\begin{document}

\title{The Linear Algebra Mapping Problem.
  Current state of linear algebra languages and libraries.}

\maketitle

\begin{abstract}
  We observe a disconnect between the developers and the end users of linear algebra libraries.
  On the one hand, the numerical linear algebra and the high-performance communities invest significant effort in the
  development and optimization of highly sophisticated numerical kernels and libraries, aiming at the maximum exploitation of
  both the properties of the input matrices, and the architectural features of the target computing platform.
  On the other hand, end users are progressively less likely to go through the error-prone and time consuming process of
  directly using said libraries by writing their code in C or Fortran; instead, languages and libraries such as Matlab, Julia, Eigen and Armadillo, which offer a higher level of abstraction, are becoming more and more popular.
  Users are given the opportunity to code matrix computations with a syntax that closely resembles the mathematical
  description; it is then a compiler or an interpreter that internally maps the input program to lower level kernels, as
  provided by libraries such as BLAS and LAPACK.
  Unfortunately, our experience suggests that in terms of performance, this translation is typically vastly suboptimal.

  In this paper, we first introduce the Linear Algebra Mapping Problem, and then investigate
  how effectively
  a benchmark of test problems
  is solved by
  popular high-level
  programming languages and libraries. Specifically, we consider
  Matlab, Octave, Julia, R, C++ with Armadillo, C++ with Eigen, and Python with NumPy;
  the benchmark is meant to test both
  standard compiler optimizations such as common subexpression elimination and loop-invariant code motion,
  as well as linear algebra specific optimizations such as optimal parenthesization for a matrix product and
  kernel selection for matrices with properties.
  The aim of this study is to give concrete guidelines for the development of languages and libraries that support
  linear algebra computations.

\end{abstract}



\section{Introduction}\label{sec:introduction}
Linear algebra expressions are at the heart of countless applications and algorithms in science and engineering, such as
linear programming~\cite{straszak2015}, signal processing~\cite{ding2016}, direct and randomized matrix
inversion~\cite{bientinesi2008,Gower:2017bq}, the Kalman and the ensemble Kalman filter~\cite{Kalman:1960ii,nino2016},
image restoration~\cite{Tirer:2017uv}, stochastic Newton method~\cite{Chung:2017ws}, Tikhonov regularization~\cite{Golub:2006hl},
and minimum mean square error filtering~\cite{Kabal:2011wr}, just to name a few.
The efficient computation of such expressions is a task that requires a thorough understanding of both numerical methods and computing architectures.
To address these requirements, the numerical linear algebra community put a significant effort into the identification and development of a relatively
small set of kernels to act as building blocks towards the evaluation of said expressions.
Such kernels are tailored for many different targets, including computing platforms, matrix properties, and data types, and are often
packaged into highly sophisticated and portable libraries, such as OpenBLAS and LAPACK\@.
However, many of the application problems encountered in practice are more complex than the operations supported by those kernels,
making it necessary to break the target problem down into a sequence of kernel invocations.
The problem we consider in this article is that of computing target linear algebra expressions, such as the ones presented in
Table~\ref{tab:LAMP_examples}, from a set of available building blocks, such as the kernels offered by the BLAS/LAPACK
libraries (see Table~\ref{tab:kerns}).
We refer to this problem as the Linear Algebra Mapping Problem (LAMP).

\begin{table}
    \centering
    \begin{adjustbox}{max width=\textwidth}
        \begin{tabular}{lll}
        	\toprule
        	Application            & Expression                                                                     & Properties       \\ \midrule
        	Standard Least Squares & $b := (X^T X)^{-1} X^T y$                                                      &                  \\ \midrule
        	Rand. Matrix Inversion & $X_{k+1} := X_k + W A^T S (S^T AWA^T S)^{-1} S^T (I_n - A X_k)$                & $W$: SPD         \\ \midrule
        	                       & $K_k   := P_{k-1} H^{T} ( H_{k} P_{k-1} H_k^T + R_k )^{-1}$                    &                  \\
        	Kalman Filter          & $P_k := \left( I - K_k H_k \right) P_{k-1}$                                    & $P$: SPD         \\
        	                       & $x_{k} \; := x_{k-1} + K_k ( z_k - H_{k} x_{k-1} )$                            & $R$: SPSD        \\ \midrule
        	Signal Processing      & $x := \left( A^{-T} B^T B A^{-1} + R^T L R \right)^{-1} A^{-T} B^T B A^{-1} y$ & $L$: DI, $R$: UT \\ \bottomrule
        \end{tabular}
    \end{adjustbox}
    \caption{Exemplary target linear algebra expressions. The properties are as follows. SP(S)D: Symmetric Positive (Semi-)Definite, DI: Diagonal, UT: Upper Triangular.}
    \label{tab:LAMP_examples}
\end{table}

\begin{table}
    \small{
      \renewcommand{\arraystretch}{1.1}
      \centering
      \begin{tabular}{lll}
      	\toprule
      	Name        & Expression                  & Description              \\ \midrule
      	\texttt{DOT}   & $\alpha := x^T y$           & inner product            \\
      	\texttt{GER}   & $A := \alpha x y^T + A$     & outer product            \\
      	\texttt{TRSV}  & $L\, x = b$                & triangular linear system\\
      	\texttt{GEMM}  & $C := \alpha A B + \beta C$ & matrix-matrix product    \\
      	\texttt{POTRF} & $L L^T = A$                & Cholesky factorization   \\
      	\texttt{SYEVR} & $Q^T T Q = A$              & eigendecomposition     \\ \bottomrule
      \end{tabular}
      \caption{Exemplary linear algebra building blocks.}
			\label{tab:kerns}
    }
\end{table}

Solutions to LAMP range from entirely manual to fully automatic.
The manual approach consists in writing a program in a low-level language such as C or Fortran, and explicitly
invoking library kernels.
This process is both time consuming and error prone:
It requires users to make decisions about which properties to exploit, which kernels to use and in which order, and all of this while adhering to rather complex APIs.
Automated solutions are provided by high-level languages and libraries such as Matlab, Julia, and Armadillo,
which allow users to write programs that closely mirror the target linear algebra expressions.
It is then a compiler/interpreter that automatically identifies how to map the input program onto the available kernels.
The quality of the mapping depends on the specific
language\footnote{From now on, we use the term ``language'' loosely, without distinguishing between programming languages, libraries and frameworks.} of choice,
but in general, it will likely be significantly lower than that of a program hand-written by an expert.
However, such automatic approaches make it possible even for non-experts to quickly obtain a working program, thus boosting productivity and enabling experimentation.
Furthermore, high-level languages give users the opportunity to partially influence how expressions are evaluated, for example by using parenthesization.
Some languages even allow for a hybrid approach, offering not only a high-level interface, but also more or less direct access to the underlying BLAS/LAPACK kernels.

The objective of this article is threefold:
First, we introduce LAMP, a term that attempts to unify a number of problems related to the efficient computation of linear algebra expressions.
Second, we assess the capabilities of current state-of-the-art languages that solve instances of LAMP\@.
The assessment is carried out by a set of minimal tests, each exposing one single optimization.
Our intention is \emph{not} to compare tools with one another, but to help users and developers understand the capabilities of each individual language.
Third, we aim to provide guidelines for the development and improvement of such languages by introducing a benchmark of high-level linear algebra optimizations.

The organization of the article follows.
In Sec.~\ref{sec:LAMP}, we define LAMP and discuss its computational complexity.
In Sec.~\ref{sec:related_work} we survey the landscape of languages, libraries, frameworks, and tools that
solve different instances of LAMP\@.
In Sec.~\ref{sec:experiments}, we introduce a benchmark of linear algebra expressions, and use it to evaluate the extent
to which high-level programming languages incorporate optimizations that play a significant role in the solution of LAMP\@.
Finally, in Sec.~\ref{sec:conclusions} we summarize our contributions and discuss ways of expanding this study.


\section{The Linear Algebra Mapping Problem}
\label{sec:LAMP}

In its most general form, LAMP is defined as follows.
Given a linear algebra expression $\mathcal{L}$, a set of instructions $\mathcal{I}$, and a cost function $\mathcal{C}$, LAMP 
consists in constructing a program $\mathcal{P}$, using the instructions from $\mathcal{I}$, that computes $\mathcal{L}$ and minimizes the cost $\mathcal{C}(\mathcal{P})$.
%
%
Depending on the specific choice of $\mathcal{L}$, $\mathcal{I}$, and $\mathcal{C}$, one will recognize that many
different, seemingly unrelated, problems are all instances of LAMP. 
A few examples follow.
\begin{itemize}
  \item When $\mathcal{L}$ is the matrix-matrix product $C := AB + C$ with variable operand sizes, $\mathcal{I}$ is the set of machine instructions, and $\mathcal{C}$ is the execution time, the problem reduces to the development of the high-performance \texttt{GEMM} kernel. This problem is central to many high-performance linear algebra libraries \cite{blas3}, and significant effort is put both into manual solutions such as GotoBLAS \cite{gotoblas2008}, OpenBLAS \cite{openblas2012} and BLASFEO \cite{Frison:2017vp}, as well as with auto-tuned libraries such as ATLAS \cite{atlas1998}.
  \item When $\mathcal{L}$ consists of a matrix product $X := M_1 M_2 \cdots M_k$, the only available instruction in $\mathcal{I}$ is the matrix product $C := AB$, and the cost function counts the number of floating point operations, LAMP reduces to the matrix chain problem \cite{cormen1990}. Several variants of this problem have been studied, including finding solutions for parallel systems \cite{lee2003} and GPUs \cite{nishida2011}.
  \item When $\mathcal{L}$ contains small-scale, memory bound problems, and $\mathcal{I}$ consists of scalar and vectorized instructions, LAMP covers the domain of code generators such as BTO BLAS \cite{Siek:2008ij}, which aims to minimize the number of memory accesses, as well as LGen \cite{spampinato2016} and SLinGen \cite{Spampinato:2018tz}, which instead minimize execution time.
  \item When $\mathcal{L}$ consists of BLAS-like operations, such as matrix inversion, least-squares problems, and the derivative of matrix factorizations \cite{smith1995, giles2008}, $\mathcal{I}$ contains BLAS/LAPACK kernels, and $\mathcal{C}$ is a performance metric, LAMP    captures problems solved by the FLAME methodology \cite{bientinesi2008, fabregat-traver2014}.
  \item When $\mathcal{L}$ is made up of matrix expressions as those shown in Table~\ref{tab:LAMP_examples}, $\mathcal{I}$ contains kernels as those shown in Table~\ref{tab:kerns}, and the cost is execution time, LAMP describes the problem that languages such as Matlab aim to solve. This class of LAMP instances is the main focus of this article.
\end{itemize}

While execution time is the most commonly used performance metric, all practical solutions to LAMP also have to fulfill
requirements regarding numerical stability. This means that in practice the cost function is
a multi-level metric, e.g., a tuple in which the first entry is a measure of numerical stability, and following ones are
performance metrics  such as execution time and data movement.

\subsection{Complexity of LAMP}


\changed{As evinced by the large number of languages and libraries that solve LAMP, finding a potentially suboptimal solution is easy. However, as we show in this section, finding the optimal solution is difficult for many variants of LAMP. Specifically,} any variant of LAMP that makes it possible to have common subexpressions is at least NP-complete.
Our proof hinges on the NP-completeness of the Optimal Common Subexpression Elimination problem (OCSE), since the optimal solution of LAMP requires the solution of OCSE.



\begin{definition}[Optimal Common Subexpression Elimination]
Let $D$ be a set, $\bullet: D \times D \rightarrow D$ be an associative-commutative operator, and $A$ a finite set of variables over $D$.
Consider
(i) a collection of equations $x_k = a_1 \bullet \ldots \bullet a_l$,
with $a_1, \ldots, a_l \in A$, and  $k = 1,\ldots,n$, where each variable appears at most once per equation, and
(ii) a positive integer $\Omega$.
Is it possible to find a sequence of assignments
$u_i = s_i \bullet t_i$, with $i = 1,\ldots,\omega$ and $\omega \leq \Omega$,
where $s_i$ and $t_i$ are either an element of $A$ or $u_j$ with $j < i$,
such that for all $k$ there exists a $u_i$ which equals $x_k$?
\end{definition}
Intuitively, given a set of assignments that contain common subexpressions, the problem consists in computing the
assignments with as few operations as possible. An instance of OCSE (left) and its solution (right) are given below:
\begin{align*}
 A       & = \{a_1, a_2, a_3, a_4\}      &    u_1 & = a_1 \bullet a_2 = x_1 \\
x_1      & = a_1 \bullet a_2             &    u_2 & = a_2 \bullet a_3            \\
x_2      & = a_1 \bullet a_2 \bullet a_3 &    u_3 & = a_1 \bullet u_2 = x_2 \\
x_3      & = a_2 \bullet a_3 \bullet a_4 &    u_4 & = a_4 \bullet u_2 = x_3 \\
\Omega   & = 4                           &        & 
\end{align*}
This example contains two common subexpressions: $a_1 \bullet a_2$ (which appears in $x_1$ and $x_2$), and $a_2 \bullet
a_3$ (which appear in $x_2$ and $x_3$). Since in $x_2$ they overlap, it is not possible to make use of them both. In
this case, using either one leads to a solution, but in general the difficulty of OCSE lies in deciding which common subexpressions to use to minimize the number of assignments $u_i$.
Since the definition of OCSE only requires one associative-commutative binary operator, the problem arises in many areas: The set $D$ can be the set of integers, real or complex numbers, but also vectors or matrices. The operator can either be addition or multiplication, with the exception of matrix multiplication, as it is not commutative.

We prove that OCSE is NP-complete by reduction from Ensemble Computation (EC) \cite{garey2002}, which is known to be NP-complete. By showing that for every instance of EC there is an equivalent instance of OCSE, we show that OCSE is at least as difficult as EC. The definition of EC is provided below.

\begin{definition}[Ensemble Computation]
  Consider (i) a collection $C = \{ C_k \subseteq A \mid k = 1,\ldots,n \}$ of subsets of a finite set $A$, and (ii) a positive integer $\Omega$. Is there a sequence $u_i = s_i \cup t_i$ for $i = 1,\ldots,\omega$, $\omega \leq \Omega$, where $s_i$ and $t_i$ are either $\{a\}$ for some $a \in A$, or $u_j$ for some $j < i$ and $s_i \cap t_i = \varnothing$, such that for all $C_k \in C$ there is a $u_i = C_k$?
\end{definition}
The idea of EC is to construct a collection of subsets $C_k$ of a set $A$ with as few binary unions as possible. For those unions, one either has to use singleton sets $\{a\}$ with $a \in A$, or intermediate results from previous unions. The similarity to OCSE lies in the challenge to optimally make use of subsets that the different $C_k$ have in common.

The NP-completeness of OCSE is demonstrated in two steps: First, we show that OCSE is in NP by showing that its solutions
can be verified in polynomial time. Then, we show that is possible to reduce EC to OCSE in polynomial time.

\begin{proof}
{\sc Verification:} A solution to OCSE can be verified in polynomial time
by traversing the sequence $u_i = s_i \bullet t_i$, $i = 1,\ldots,\omega$, collecting the sets of all variables that contribute to each $u_i$, and comparing those sets with the
right-hand sides of the $n$ input equations $x_k = \dots$ with $k=1,\dots,n$.

{\sc Reduction.} For each instance of EC, an equivalent instance of OCSE is obtained as follows.
  For each $C_k$, an in input equation is constructed as $x_k = a_1 \bullet \ldots \bullet a_l$ with all $a_1, \ldots, a_l \in C_k$.
  In the solution, the sets $\{a_i\}$ are substituted with the corresponding variables $a_i$, and the unions $u_i = s_i \cup t_i$ with operations $u_i = s_i \bullet t_i$.
\end{proof}


%

We conclude with the EC instance (left) and its solution (right) that correspond to the OCSE instance shown above:
\begin{align*}
A &= \{a_1, a_2, a_3, a_4\}             & u_1 &= \{a_1\} \cup \{ a_2 \} = C_1 \\
C &= \{\{a_1, a_2\}, \{a_1, a_2, a_3\}, \{a_2, a_3, a_4\} \}
                                        & u_2 &= \{a_2\} \cup \{ a_3 \} \\
\Omega &= 4                             & u_3 &= \{a_1\} \cup u_2 = C_2 \\
&                                       & u_4 &= \{a_4\} \cup u_2 = C_3
\end{align*}

\changed{

Common subexpressions are not the only reason why variants of LAMP can be at least NP-complete. Since LAMP bears similarities to code generation for scalar code, results carry over.
For example, in practice the amount of available memory is limited. Thus, it could be important to identify if a given
sequence of kernels can be computed with a certain amount of memory, under the assumption that kernels can be reordered
as long as the data-flow dependencies are satisfied. It is possible to show that this problem is NP-complete by
reduction from Register Sufficiency \cite[App.~A11.1]{garey2002}.

Register Sufficiency is the problem of identifying whether a given program which is described in terms of a dependency
graph can be computed with at most $k$ registers. The input to the Register Sufficiency problem is a directed acyclic
graph that represent the data-flow dependencies between the instructions of the program. The problem then consists in
finding an ordering of the nodes (the instructions) that satisfies the dependencies which can be computed with at most
$k$ registers.

%
%

Similarly, one could also reduce the Register Allocation problem \cite{chaitin1981} to LAMP.\footnote{Notice that
  whether or not Register Allocation is NP-complete depends on the exact definition of the problem; the details are discussed in \cite{bouchez2006}.}
Given a fixed sequence of
instructions, Register Allocation is the problem of finding an optimal assignment of variables to registers that
minimizes the cost of loads and stores. The role of the registers in the Register Allocation problem is played by the
cache in LAMP.

}

\section{Related Work}\label{sec:related_work}
A considerable number of languages, libraries, frameworks, and tools are available for the solution of different instances  of LAMP\@.
In this section, we highlight those that support a high-level notation for linear algebra expressions and provide
some degree of automation in the construction of efficient solutions.
Furthermore, we survey a number of kernel libraries, which offer the necessary building blocks for higher-level LAMP ``solvers''.

\subsection{Languages}\label{subsec:languages}
Several languages and development environments have been created for scientific computations.
Matlab~\cite{matlab} is a popular language with extensions (toolboxes) for many scientific domains.
GNU Octave~\cite{gnuoctave} is open source software which supports similar functionality and syntax to Matlab.
Julia~\cite{julia2017} is a rapidly emerging language;
it features just in time compilation, and uses a hierarchical type system paired with multiple dispatch.
While the main focus of the R language~\cite{rlanguage} is on statistics, it also supports linear algebra computations.
Further examples of computer algebra systems that natively support linear algebra are Mathematica~\cite{Mathematica} and Maple~\cite{maple}.
All these languages provide mechanisms that help solve certain instances of LAMP\@.

\subsection{Libraries}\label{subsec:libraries}
For virtually every established high-level programming langua\-ge, libraries for linear algebra computations exist.
The idea is usually to offer a domain-specific language for linear algebra within the host language, usually by adding classes for matrices and vectors, by overloading operators and in the case of C++, by expression templates.
Expression template libraries for C++ include: Eigen~\cite{eigenweb}, Blaze~\cite{iglberger2012a}, Armadillo~\cite{sanderson2010}, HASEM~\cite{hasem}, MTL4~\cite{mtl42007}, uBLAS~\cite{ublas2006}, and blitz++~\cite{blitz1998}.
They offer a compromise between ease of use and performance.
Similar libraries exist for many other languages; examples include NumPy~\cite{oliphant2006numpy} for Python and the Apache Commons Mathematics Library~\cite{apachecommonsmath} and ND4j~\cite{nd4j2016} for Java.
By virtue of these libraries, users of general purpose programming languages are exposed to some of the LAMP solving functionality that is available in linear algebra targeted languages.

\subsection{Tools and Algorithms}\label{subsec:tools-and-algorithms}
The Transfor program~\cite{gomez1998} is likely the first translator of linear algebra expressions (written in Maple) into BLAS kernels.
More recently, several other solutions to different variants of LAMP have been developed.
CLAK~\cite{fabregat-traver2013a} and its successor, Linnea~\cite{linnea2019},
are tools that receive a linear algebra expression as input and produce as output a sequence of calls to BLAS and LAPACK that compute the input expression.
The Formal Linear Algebra Methods Environment (FLAME)~\cite{Gunnels:2001gi,Bientinesi:2005hu} is a methodology for the
derivation of algorithmic variants for BLAS-like operations and for equations such as triangular Sylvester and Lyapunov;
Cl1ck~\cite{FabregatTraver:2011km,FabregatTraver:2011gu} is an automated implementation of the FLAME methodology.
The goal of BTO BLAS~\cite{Siek:2008ij} is to generate C code for bandwidth bound operations, such as fused matrix-vector operations.
DxTer~\cite{marker2013} uses domain knowledge to optimize programs represented as dataflow graphs.
LGen~\cite{spampinato2016} targets basic linear algebra operations for small operand sizes, a regime in which BLAS and LAPACK do not perform very well, by directly generating vectorized C code.
SLinGen~\cite{Spampinato:2018tz} combines Cl1ck and LGen to generate code for more complex small-scale problems.
The generalized matrix chain algorithm~\cite{barthels2018cgo} is an extension of the standard matrix chain algorithm~\cite{cormen1990};
it finds the optimal solution (in terms of FLOPs) for matrix chains with operands that can be transposed or inverted, and considers matrix properties.
LINVIEW~\cite{linview2014} introduces techniques for incremental view maintenance of linear algebra.

\subsection{Kernel Libraries}\label{sec:related_work:kernels}
Kernels are highly optimized routines that perform relatively simple operations, and that allow more complex algorithms
to be structured in a layered fashion.
In numerical linear algebra, the BLAS specification was introduced to standardize vector~\cite{blas1}, matrix-vector~\cite{blas2} and matrix-matrix~\cite{blas3} operations,
and to assist the development of highly optimized libraries.
Several libraries offer optimized BLAS implementations, including GotoBLAS~\cite{gotoblas2008}, OpenBLAS~\cite{openblas2012}, BLIS~\cite{VanZee:2015go}, BLASFEO~\cite{Frison:2017vp}, clBLAST~\cite{clblast2018}, and LIBXSMM~\cite{libxsmm2016}.

Built on top of BLAS, kernels for more complex operations (e.g., solvers for linear systems, least-squares problems, and eigenproblems) are offered in 
LAPACK~\cite{lapack1990} 
libflame~\cite{libflame2011}, and RELAPACK~\cite{relapack2016}.
Proprietary kernel libraries that implement a superset of BLAS and LAPACK include
Intel MKL~\cite{mkldoc}, Nvidia cuBLAS~\cite{cublas}, IBM ESSL~\cite{essl} and the Apple Accelerate Framework~\cite{accelerate}.

Similar libraries exist for sparse  computations, including
PSBLAS~\cite{Filippone:2000hk}, \\
clSparse~\cite{clsparse2016}, HSL (formerly the Harwell Subroutine Library)~\cite{Hopper:1973wu}, and PETSc~\cite{petsc2018}.

\section{Evaluation of Programming Languages}
\label{sec:experiments}

Several programming languages make it possible for users to input linear algebra expressions almost as if they were writing them on a blackboard.
For instance, in Matlab/Octave, Armadillo, and Julia,
the assignment $C := AB^{T} + BA^{T}$ can be written as
\verb|C = A*B' + B*A'|,\\
\verb|C = A*trans(B) + B*trans(A)|, and
\verb|C = A*transpose(B) + B*transpose(A)|,
respectively.
When using such a level of abstraction, users relinquish control on the actual evaluation of the expressions,
effectively relying on the internal mechanisms of the language to solve LAMP.

In this section, we consider seven such
languages---Armadillo, Eigen, Julia, Matlab, NumPy (Python),
GNU Octave and R\footnote{This is by no means an exhaustive list of languages that offer a high-level API for linear algebra;
others exist (e.g., Mathematica and Maple).
In our experience, the languages considered are among the most commonly used for numerical computations and data analysis applications.}---and
introduce a benchmark to assess how efficiently they solve a number of test expressions.
These expressions were designed to be as simple as possible, while capturing, in isolation, scenarios that occur
frequently in practice and for which one specific optimization is applicable.
The results, in terms of execution time, are compared to an ``expert'' implementation, written either in the same language or in C.
This comparison aims to showcase the extent to which each language implements an optimization and is \emph{not} intended for ranking the different languages.
Ultimately, this section is meant to evaluate the quality of the solutions provided by the languages that solve LAMP, thus inspiring and guiding their development.

\subsection{Setup}\label{subsec:setup}

Our benchmark consists of 12 experiments, each one of them containing one or more test expressions, to be used as input to the languages.
In all cases, the input programs (expressions) resemble the mathematical representation as closely as possible. 
Consequently, whenever an operation is supported by both a function and an operator, the latter is preferred (e.g., for matrix
multiplication, NumPy supports both the function \verb|matmul| and the operator \verb|@|).
Furthermore, the input expressions are as compact as possible, that is, not broken into multiple assignments and without explicit parenthesization.
In addition, all matrices (input and output) are preallocated and initialized before any timing.
Finally, the operands are chosen large enough so that the individual timings are less susceptible to noise and fluctuations. 

For each experiment, we report the minimum execution time over 20 repetitions, flushing all cache memories in between each repetition.
Special measures are taken to avoid dead code elimination in both the experiments and cache flushing.
For those languages that have a garbage collector (Julia\changed{, Python,} and R),
we explicitly invoke it after cleaning the cache to reduce the chances of interference with our timings.
Furthermore, we do not concern ourselves with how much time it takes for languages to make decisions;
rather, we evaluate the quality of those decisions and assess whether or not a specific optimization is implemented.

The experiments are performed on \changed{a single core of} a Linux machine with an Intel Xeon E5-2680V3 processor, with Turbo Boost disabled.
All languages are linked to the Intel(R) Math Kernel Library 19.0, which implements a super-set of BLAS and LAPACK, and
compiled with gcc\footnote{Version 8.2.0 with optimization flag -O3}.
The versions of the languages used are the latest stable releases as of \changed{December 2020:
Armadillo 10.1.x, Eigen 3.3.8, Julia 1.5.2, Matlab 2020a, GNU Octave 5.2.0, NumPy 1.19.4 and R 4.0.3.}
The source code for the experiments is available online\footnote{\url{https://github.com/ChrisPsa/LAMP_benchmark}}.

\subsection{Mapping to Kernels}
\label{sec:methodology_explanation}
BLAS and LAPACK offer a set of kernels that are the de-facto standard building blocks for linear algebra computations.
Many optimized implementations of such operations exist, see Section~\ref{sec:related_work:kernels}.
All the aforementioned languages have access to optimized kernels via MKL.
Here we investigate the capabilities of modern linear algebra languages in mapping fundamental operations to BLAS kernel calls.

\subsubsection{Experiment \texttt{\#}1: GEMM}
\paragraph{Input}
In this first experiment, we initialize the random matrices \dimms{A}{m}{k}, \dimms{B}{k}{n}, and \dimms{C}{m}{n}, and
we input the expression $C := AB$ in each language by using the available matrix representations (objects) and the operator for matrix multiplication.
The goal is to determine whether languages compute this expression by invoking the optimized BLAS kernel \texttt{GEMM}, or via another (inferior) implementation.
The \texttt{GEMM} kernel included in the optimized BLAS libraries is an extremely sophisticated piece of code~\cite{gotoblas2008};
consequently, the difference in performance between a call to \texttt{GEMM} and to any other (suboptimal) implementation is
going to be significant and easily distinguishable by comparing the execution time with that of
an explicit call to \texttt{GEMM} implemented in C, henceforth referred to as ``reference''.

\paragraph{Results}
Table~\ref{tab:interpretation_merged} shows the execution time for each language to perform the matrix product.
The expectation is that if the timings are ``close enough'' to the reference, then it
can be inferred that the languages do rely on the \texttt{GEMM} kernel, modulo some overhead.
The timings indicate that all languages are within 15\% of the execution time of the reference, thus providing strong evidence that they all invoke the optimized \texttt{GEMM}\@.

\subsubsection{Experiment \texttt{\#}2: SYRK}
\paragraph{Input}
Since all languages successfully map to \texttt{GEMM}, in this second experiment, we initialize the random matrices
\dimms{A}{n}{k}, \dimms{C}{n}{n}, and input the expression $C := A A^{T}$, which is a special instance of \texttt{GEMM} in which matrix $B$ is substituted with $A^{T}$.
Similarly to Experiment \texttt{\#}1, we make use of the high-level abstractions offered by each language.
Although the output matrix $C$ could be computed with a call to \texttt{GEMM}, performing $2n^{2}k$ FLOPs (``Floating Point Operations''),
BLAS offers a specialized routine, \texttt{SYRK} (``SYmmetric Rank-K update''), which only performs $n^{2}k$ FLOPs.
One expects \texttt{SYRK} to complete in approximately half the execution time of \texttt{GEMM}\@.
As a reference implementation, we also performed a call to \texttt{SYRK} in C\@.

\paragraph{Results}
In Table~\ref{tab:interpretation_merged}, by comparing the timings for \texttt{SYRK} to those of \texttt{GEMM},
one can tell which languages take advantage of the specialized routine, and which do not.
Specifically, most languages have a computation time that is significantly less than that of a \texttt{GEMM} (almost half),
strongly suggesting that they make the right decision.
However, this is not the case for Eigen and R, whose computation time is equal to that of a \texttt{GEMM}\@.

\newpage
\subsubsection{Experiment \texttt{\#}3: SYR2K}
\paragraph{Input}
Since several languages are able to tell apart \texttt{SYRK} and \texttt{GEMM}, we now initialize the random matrices
\dimms{A, B}{n}{k}, \dimms{C}{n}{n}, and test the slightly more complex expression $C := A B^{T} + BA^{T}$.
This assignment could be computed by two successive calls to \texttt{GEMM}; it is however supported by the \texttt{SYR2K} kernel (``SYmmetric Rank-2K update''),
which---similarly to \texttt{SYRK}---takes advantage of the fact that the matrix $C$ is symmetric (cost: $2n^{2}k$ FLOPs).
Therefore, its execution time is expected to be approximately equal to that of a \texttt{GEMM}\@.
As a reference implementation, we performed a call to \texttt{SYR2K} in C\@.

\paragraph{Results}
In Table~\ref{tab:interpretation_merged}, by comparing the timings for \texttt{SYR2K} to those for \texttt{GEMM} and to the reference,
one observes that in all cases, \texttt{SYR2K} requires double the time of a \texttt{GEMM},
thus indicating that no language selects the specialized BLAS kernel for \texttt{SYR2K}\@.

\begin{table}
    \centering
    \begin{adjustbox}{max width=\textwidth}
    \begin{tabular}{llcccccccc}
    	\toprule
    	Name                            & Expression                              &  C   & Armadillo & Eigen & Julia & Matlab & NumPy & Octave &  R   \\ \midrule
    	\multirow{2}{*}{\texttt{GEMM}}  & \multirow{2}{*}{$C := AB$}              & 1.43 &   1.43    & 1.46  & 1.44  &  1.44  & 1.48  &  1.48  & 1.47 \\
    	                                &                                         &      &   \yes    & \yes  & \yes  &  \yes  & \yes  &  \yes  & \yes \\ \midrule
    	\multirow{2}{*}{\texttt{SYRK}}  & \multirow{2}{*}{$C := AA^{T}$}          & 0.73 &   0.74    & 1.57  & 0.76  &  0.76  & 0.78  &  0.78  & 1.51 \\
    	                                &                                         &      &   \yes    &  \no  & \yes  &  \yes  & \yes  &  \yes  & \no  \\ \midrule
    	\multirow{2}{*}{\texttt{SYR2K}} & \multirow{2}{*}{$C := AB^{T} + BA^{T}$} & 1.47 &   2.91    & 2.89  & 2.92  &  2.91  & 2.96  &   3    & 3.04 \\
    	                                &                                         &      &    \no    &  \no  &  \no  &  \no   &  \no  &  \no   & \no  \\ \bottomrule
    \end{tabular}
    \end{adjustbox}
    \caption{Experiments \texttt{\#}1-3. Timings are in seconds. By comparing the execution time of each language with
      that of hand-written C code, one can deduce whether or not a language makes use of the most appropriate BLAS kernel for the evaluation of each expression.}
    \label{tab:interpretation_merged}
\end{table}

\subsubsection{Experiment \texttt{\#}4: Update of C}
As specified in the BLAS interface~\cite{blas3}, the kernels \texttt{GEMM}, \texttt{SYRK} and \texttt{SYR2K} offer the option of updating the matrix $C$\@.
The full definition of \texttt{GEMM} is $C := \alpha A B + \beta C$, where $\alpha$, $\beta$ are scalars and $A$, $B$ and $C$ are matrices.
Therefore, expressions such as $C := AB + C$ can be computed using one single call to \texttt{GEMM}, without the need for intermediate storage for $AB$.
This functionality, which is also supported by \texttt{SYRK} and \texttt{SYR2K}, increases the overall performance and reduces the size of temporary storage;
however, the computational cost for the addition of two matrices of size \dims{n}{n} is \order{n^2},
and for mid- and large-sized matrices this will be dwarfed by the \order{n^3} cost for the multiplication.
On the contrary, the smaller the problem size, the more significant the contribution of the addition to the overall computation time.
For completeness, we investigate whether or not the languages require a separate matrix addition when given such expressions as input.

\paragraph{Input}
We used the expressions in Table~\ref{tab:updatec} as input, where the matrices have the same sizes as in the three experiments above.
To test if the languages require an extra addition, we also measured the time it takes for a similarly sized matrix addition in each language.

\paragraph{Results}
Timings (see Table~\ref{tab:app_updatec}) suggest that in all cases the expression is computed as two steps, a matrix multiplication followed by a matrix addition.
The only exception is the ``$\pluseq$'' operator overload in Armadillo for \texttt{GEMM}.

\begin{table}
    \centering
    \begin{adjustbox}{max width=\textwidth}
    \begin{tabular}{l@{\hspace*{2pt}}r@{\hspace*{2pt}}l@{\hspace*{2pt}}l@{\hspace*{2pt}}lccccccc}
    	\toprule
    	\multicolumn{5}{l}{Expression}                    & Armadillo & Eigen & Julia & Matlab & NumPy & Octave &  R   \\ \midrule
    	$C$       &      $:=$ & $AB$     &     &          &   \yes    & \yes  & \yes  &  \yes  & \yes  &  \yes  & \yes \\
    	$C$       &      $:=$ & $AB$     & $+$ & $C$      &    \no    &  \no  &  \no  &  \no   &  \no  &  \no   & \no  \\
    	$C$       & $\pluseq$ & $AB$     &     &          &   \yes    &  \no  &  \no  &  \na   &  \no  &  \no   & \na  \\ \midrule
    	$C$       &      $:=$ & $AA^{T}$ &     &          &   \yes    &       & \yes  &  \yes  & \yes  &  \yes  &      \\
    	$C$       &      $:=$ & $AA^{T}$ & $+$ & $C$      &    \no    &       &  \no  &  \no   &  \no  &  \no   &      \\
    	$C$       & $\pluseq$ & $AA^{T}$ &     &          &    \no    &       &  \no  &  \na   &  \no  &  \no   &      \\ \midrule
    \end{tabular}
    \end{adjustbox}
    \caption{Experiment \texttt{\#}4: Update of C. With the exception of the $\pluseq$ operator overload in Armadillo
      for \texttt{GEMM}, no language maps to one single kernel call which includes the update to C.}
    \label{tab:updatec}
\end{table}

\subsection{Linear Systems}
Although matrix inversion is an extremely common operator in linear algebra expressions, only selected applications actually require the explicit inversion of a matrix.
In the vast majority of cases, the inversion can (and should) be avoided by solving a linear system, gaining both in speed and numerical stability~\cite[p. 260]{higham2002}.
However, we observed that it is extremely common for inexperienced users to blindly translate the mathematical
representation into code, resulting in the expressions such as $(AB + C)^{-1}Y$ being coded in Matlab as \verb|inv(AB + C)*Y|, instead of the recommended \verb|(AB+C)\Y|.

As shown in Table~\ref{tab:examples}, most languages provide a special function (or operator) for solving linear systems of the form $Ax=B$ (or $xA=B$),
where \dimms{A}{n}{n} is a matrix and $x$ and $B$ are either a vector of size \dims{n}{1}, or a matrix (multiple right-hand sides) of size \dims{n}{m}.
These functions are usually quite sophisticated and try to determine certain properties of $A$, so that the most
suitable (in terms of data structure, speed, and accuracy) factorization can be used.
The extent to which languages can determine those properties will be further investigated in Section~\ref{sec:properties}.

\subsubsection{Experiment \texttt{\#}5: Explicit Inversion}

\paragraph{Input}
We examine how languages handle the inverse operator;
specifically, we aim to determine whether or not languages avoid (if possible) the explicit computation of a matrix inverse.
The input to each language is the expression \verb|inv(A)*b|, where $A \in$ \dims{n}{n} matrix and $b$ is a vector of size \dims{n}{1}.
We compared the execution time to that of the expression \verb|A\b| or \verb|solve(A, b)|.

\paragraph{Results}
The timings in Table~\ref{tab:explicit_inversion} indicate that Armadillo is the only language that substitutes the \verb|inv| function
with a \verb|solve| (or ``\verb|\|'' operator).
It should be noted, however, that most languages provide warnings either during runtime or in their documentation that
using explicit inversion should be avoided whenever possible, in favor of their solve functions.
The automatic replacement of the \verb|inv| function with a solve is a rather bold decision that alters the semantics of the input expression.
For this reason, it is questionable whether or not this optimization is reasonable.
In light of the extremely common misuse of inversion in application codes, we feel that the replacement is at least partly justified.
Indeed, our recommendation is that languages automatically map calls to \verb|inv| operator to a linear system (whenever possible),
and that the actual matrix inversion is offered by less convenient functions such as \verb|explicit_inverse|.

\begin{table}[h]
    \centering
    \begin{adjustbox}{max width=\textwidth}
    \begin{tabular}{lccccccc}
    	\toprule
    	Operation       & Armadillo & Eigen & Julia & Matlab & NumPy & Octave &  R   \\ \midrule
    	\verb|inv(A)*b| &   0.68    & 2.26  & 1.74  &  1.79  & 2.26  &  1.87  & 2.25 \\
    	\verb|A\b|      &   0.68    & 0.69  & 0.68  &  0.74  & 0.71  &  0.77  & 0.73 \\ \bottomrule
    \end{tabular}
    \end{adjustbox}
    \caption{Experiment \texttt{\#}5: Explicit Inversion. Armadillo is the only language that replaces the explicit inversion of a matrix with the solution of a linear system.}
    \label{tab:explicit_inversion}
\end{table}

\begin{savenotes}
\begin{table}[H]
    \centering
    \begin{adjustbox}{max width=\textwidth}
    \begin{tabular}{ll}
    	\toprule
    	Name          & Solve linear system                                                                                                                                      \\ \toprule
    	Armadillo     & \verb|solve(A, B)|                                                                                                                                       \\
    	Eigen         & \na\footnote{Eigen does not provide a general case solve function. One must explicitly factorize and then solve a linear system with specific methods.}  \\
    	Julia         & \verb|A\B|                                                                                                                                               \\
    	Matlab/Octave & \verb|A\B|                                                                                                                                               \\
    	NumPy         & \verb|np.linalg.solve(A, B)|                                                                                                                             \\
    	R             & \verb|solve(A, B)|                                                                                                                                       \\ \bottomrule
    \end{tabular}
    \end{adjustbox}
    \caption{Functions and operands for solving linear systems.}
    \label{tab:examples}
\end{table}
\end{savenotes}

\subsection{Matrix Chains}\label{subsec:matrix-chains}
Because of associativity, a chain of matrix products (a ``matrix chain''), can be computed in many different ways,
each identified by a specific parenthesization.\footnote{The number of different parenthesizations for a chain of length $n$ is given by the Catalan number $C_{n-1} = \frac{(2n)!}{(n+1)!n!}$.}
Depending on the size of the matrices in the chain, different parenthesizations lead to vastly different execution times
and temporary storage requirements.
The problem of determining the best parenthesization, in terms of number of floating point operations,
is commonly referred to as the Matrix Chain Problem (MCP)~\cite{godbole1973,hu1982,barthels2018cgo}.
In practice, different parenthesizations may also lead to different results because floating point arithmetic is not associative.
However, since no convention for evaluating a product of matrices exists, languages can evaluate a chain in any order.

\subsubsection{Experiment \texttt{\#}6: Optimal Parenthesization}

\begin{figure}[ht]
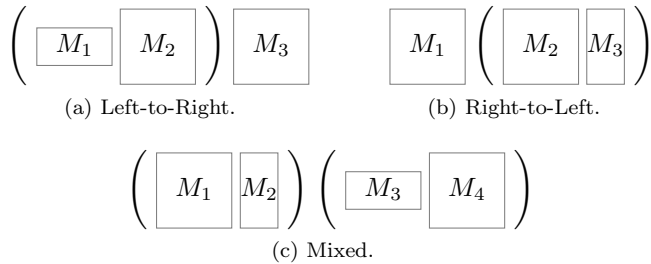

    \centering
     \subfloat[Left-to-Right.]{
        \begin{tabular}{r@{}l@{}l@{}}
            \Bigg( $\drawmatrix[height=.5]{M_1}$ &  \matprod $\drawmatrix{M_2}$ \Bigg)&  \matprod $\drawmatrix{M_3}$\\
        \end{tabular}
        \label{fig:mc2}
    } \hspace{0.5cm}
    \subfloat[Right-to-Left.]{
        \begin{tabular}{l@{}r@{}l@{}}
            $\drawmatrix{M_1}$ & \matprod\Bigg( $\drawmatrix{M_2}$ & \matprod $\drawmatrix[width=.5]{M_3}$ \Bigg)\\
        \end{tabular}
        \label{fig:mc1}
    }
\\
    \subfloat[Mixed.]{
        \begin{tabular}{r@{}l@{}r@{}l@{}}
            \Bigg( $\drawmatrix{M_1}$ & \matprod $\drawmatrix[width=.5]{M_2}$ \Bigg) & \matprod \Bigg( $\drawmatrix[height=.5]{M_3}$ & \matprod $\drawmatrix{M_4}$ \Bigg) \\
        \end{tabular}
        \label{fig:mc3}
    }
    \caption[]{Visual representation of the input expressions for Experiment \texttt{\#6}: Optimal Parenthesization.}
    \label{fig:matrix_chain_setup}
\end{figure}

\paragraph{Input}
This experiment consists of three different matrix chains.
Each of these is given as input to the languages as a single statement, and without any parenthesization.
As Fig.~\ref{fig:matrix_chain_setup} shows, the sizes of the matrices are chosen so that the optimal evaluation order
is (a) left to right, (b) right to left, and (c) a combination of the two.
The goal is to determine whether or not languages adjust the evaluation order to minimize the number of FLOPs.
For each language, the execution time is compared to that obtained for the same chain, but explicitly guided by the optimal parenthesization.

\paragraph{Results}
Table~\ref{tab:matrix_chain} indicates that most languages evaluate the chain from left to right, without considering the MCP.
Armadillo is the only language that partially solves the problem, by checking whether to evaluate from left or right;
however, it does not properly handle the mixed case.
It should be noted that NumPy offers a function called \verb|multi_dot|, which solves the MCP using dynamic programming, although the user has to explicitly invoke it.
Furthermore, several third-party developed packages that consider the MCP exist in Eigen, Julia, Matlab, and R.

\begin{table}
    \centering
    \begin{adjustbox}{max width=\textwidth}
    \begin{tabular}{lccccccc}
    	\toprule
    	Evaluation Sequence  & Armadillo & Eigen & Julia & Matlab & NumPy & Octave &  R   \\ \midrule
    	LtR no parenthesis   &   0.59    & 0.59  & 0.58  &  0.59  & 0.60  &  0.59  & 0.62 \\
    	LtR parenthesis      &   0.59    & 0.59  & 0.58  &  0.59  & 0.60  &  0.59  & 0.62 \\
    	Left-to-Right        &   \yes    & \yes  & \yes  &  \yes  & \yes  &  \yes  & \yes \\ \midrule
    	RtL no parenthesis   &   0.60    & 1.75  & 1.74  &  1.74  & 1.78  &  1.78  & 1.77 \\
    	RtL parenthesis      &   0.60    & 0.60  & 0.59  &  0.59  & 0.59  &  0.60  & 0.63 \\
    	Right-to-Left        &   \yes    &  \no  &  \no  &  \no   &  \no  &  \no   & \no  \\ \midrule
    	Mixed no parenthesis &   2.03    & 2.07  & 2.05  &  2.05  & 2.08  &  2.10  & 2.07 \\
    	Mixed parenthesis    &   0.89    & 0.92  & 0.90  &  0.90  & 0.91  &  0.92  & 0.92 \\
    	Mixed                &    \no    &  \no  &  \no  &  \no   &  \no  &  \no   & \no  \\ \bottomrule
    \end{tabular}
    \end{adjustbox}
    \caption{Experiment \texttt{\#}6: Optimal Parenthesization.
    Armadillo is the only language that incorporates a (partial) solution to the matrix chain problem.}
    \label{tab:matrix_chain}
\end{table}

\subsection{Properties}
\label{sec:properties}
BLAS \& LAPACK offer specialized kernels for specific types of operands (e.g., SPD, Symmetric, Triangular, Banded matrices\dots).
We test the matrix multiplication and the solution of a linear system, and investigate if high-level languages make use
of those kernels without the explicit help of the user.
To this end, we purposely do not use annotations about properties either in the matrix construction or in the computation.
One could---correctly---argue that in this experiment languages are not used to their best potential.
The rationale for not specifying properties is threefold:
First, we aim to capture the scenario in which non-proficient users are not aware of properties, or do not know how to exploit them.
Second, matrices can have many different origins, e.g. the explicit construction with a specialized function,
or the evaluation of an expression, and it is not guaranteed that the resulting matrices are always correctly annotated.
Finally, there are cases where properties are only known at runtime.


\begin{table}
    \begin{adjustbox}{max width=\textwidth}
    \begin{tabular}{llccccccc}
    	\toprule
    	     \multicolumn{2}{c}{Experiment}     &         &       &       &         &         &        &     \\
    	\cmidrule(r){1-2} Operation & Property  &  Armadillo   & Eigen & Julia & Matlab  & NumPy  &Octave  &   R  \\ \midrule
    	Multiplication              & Triangular&   \no   &  \no  &  \no  &   \no   &  \no    &  \no   &  \no \\
    	                            & Diagonal  &   \no   &  \no  &  \no  &   \no   &  \no    &  \no   &  \no \\\midrule
    	Linear System               & Symmetric &   \no   &  \na  &  \no  &   \no   &  \no    &  \no   &  \no \\
    	                            & SPD       &  \yes   &  \na  &  \no  &  \yes   &  \no    & \yes   &  \no \\
    	                            & Triangular&  \yes   &  \na  & \yes  &  \yes   &  \no    & \yes   &  \no \\
    	                            & Diagonal  & \changed{\symtwo} &  \na  & \yes  & \symone &  \no    & \symone&  \no \\
    	\bottomrule
    \end{tabular}
    \end{adjustbox}
    \caption{Experiments \texttt{\#}7--8. The {\symone} indicates that the solver for triangular linear systems is used
      instead of a more efficient algorithm for diagonal systems. \changed{The {\symtwo} indicates that a band matrix solver is used.}}
    \label{tab:solve_mult}
\end{table}

\newpage
\subsubsection{Experiment \texttt{\#}7: Multiplication}

\paragraph{Input}
For the multiplication of matrices with properties, we examine two cases: Triangular and Diagonal.
In the Triangular case, we input the expression $B := AB$,
where $A$ and $B$ are a Lower Triangular and a Full matrix, respectively.
We compare with a C program that explicitly invokes the BLAS kernel \texttt{TRMM}, which performs half of the FLOPs of a \texttt{GEMM} ($n^3$ vs.~$2 n^3$).
For the Diagonal case, we input the expression $C := AB$, where $A$ and $B$ are a Diagonal and a Full matrix, respectively.
Since BLAS offers no kernel for this operation, as a C language reference, one could use a loop to scale each row of $B$ individually, via the kernel \texttt{SCAL}.

\paragraph{Results}

None of the languages examined use specialized kernels or methods to perform a multiplication between a Triangular/Diagonal and a Full matrix.
It should be noted that Julia enables the user to easily annotate matrices with types that encode certain matrix properties such as Lower/Upper Triangular, Symmetric, Diagonal and more.
Specifically, Julia uses multiple dispatch~\cite{julia2017}, a technique with which it can separately define the multiplication operation for the pairs Triangular-Full and Diagonal-Full, and achieve high performance for these operations by mapping to the most appropriate kernels.
Similarly, if a matrix is created with the \texttt{diag} function, Octave stores the information that the matrix is
Diagonal, and then uses the annotation to select an efficient multiplication strategy.
However, in both cases, the effectiveness of those mechanisms depends heavily on the method used to create or initialize the matrices, as well as the ability of those languages to propagate those properties across intermediate computations.

\subsubsection{Experiment \texttt{\#}8: Properties in Linear Systems}
\paragraph{Input}
We created general matrices to satisfy the properties shown in the ``Property'' column of Table~\ref{tab:solve_mult}, and used the expressions in Table \ref{tab:examples}.
We measured the execution time and compared it to the C implementation.
The results are displayed in Table~\ref{tab:solve_mult}, while the execution time results are in Table~\ref{tab:app:solve_mult}.

\paragraph{Results}
Armadillo performs the optimal Cholesky factorization for the SPD case and forward substitution for the Triangular case;
\changed{For the Diagonal case, a band matrix solver is used.}
Eigen does not participate in this experiment, as it requires the user to explicitly specify the type of factorization to perform on the input operand before solving the linear system.
Julia recognizes the Triangular and Diagonal properties and performs forward substitution and vector scaling respectively for those cases.
However, it does not make use of the Cholesky factorization for SPD matrices nor of the Bunch–Kaufman decomposition for symmetric matrices.
Matlab is known\footnote{https://uk.mathworks.com/help/matlab/ref/mldivide.html} to have an elaborate decision tree when
solving a linear system to select the most suitable factorization based on the properties of the operands.
Those properties are detected during runtime either by examining the contents of the matrices or by trial-and-error.
Indeed, both Matlab and Octave take advantage of the SPD and the Triangular case;
the Diagonal case is treated like Triangular.
Finally, timings in Table~\ref{tab:app:solve_mult} suggest that Python and R use the general purpose LU factorization for all cases.


\subsection{Common Subexpression Elimination}
A common feature of modern compilers, at least when it comes to scalar computations, is Common Subex\-pres\-sion Elimination (CSE).
Compilers perform data flow analysis to detect subexpressions that evaluate to the same value and assess whether or not it is beneficial to compute them only once and substitute them with a temporary value in all subsequent instances.
In Sec.~\ref{sec:LAMP}, we proved that the optimal selection of common subexpressions is an NP-complete problem.

In light of the increased computational cost of matrix operations compared to scalar operations, it is mostly beneficial to detect and eliminate common subex\-pres\-sions within linear algebra expressions.
Consider for example the expression which occurs in the Stochastic Newton equations~\cite{Chung:2017ws}: $B_1 :=
\frac{1}{\lambda_1} (I_n - A^T W_1 (\lambda_1 I_l + W_1^T A A^T W_1)^{-1} W_1^T A  )$, where \dimms{A}{m}{n} and \dimms{W_1}{m}{l} are general matrices.
The term $A^{T}W_{1}$ appears a total of four times in its original and transposed form, and can be factored out and
computed only once, saving $6nml$ FLOPs.
However, when dealing with matrix expressions, the elimination of common subexpressions might be counter-productive, as the following example illustrates:
Consider the expression $A^{-T} B^T B A^{-1} y$, where $A, B \in$ \dims{n}{n} and $y \in$ \dims{n}{1}, which appears in signal processing~\cite{ding2016}.
First, the expression has to be changed to the form $(B A^{-1})^{T} B A^{-1}$ for the common subexpression to appear.
Then, one might be tempted to factor out $K := B A^{-1}$, solve it and then proceed to compute $K^{T}Ky$. The cost
  of this strategy is $\frac{8 n^{3}}{3} + \frac{7 n^{2}}{2}$.
By contrast, the optimal solution is to evaluate the initial expression from right to left, for a cost of $\frac{4 n^{3}}{3} + 7 n^{2}$.
Furthermore, in addition to the computational cost, the decision to eliminate a common subexpression has to take into account the memory overhead of temporary matrices, which might represent a hard constraint, especially for architectures with limited memory.

\subsubsection{Experiment \texttt{\#}9: Common Subexpressions}

\paragraph{Input}
To identify if any of the languages performs CSE, we create the random matrices \dimms{A, B}{n}{n} and use the two expressions in Table~\ref{tab:CSE} as input.
In the ``naive'' column, the product $AB$ appears twice; in the ``recommended'' column, the product is factored out with the help of a temporary variable.

\paragraph{Results}
By comparing the execution time for the two experiments in Table~\ref{tab:app:cse} in Appendix~\ref{app:timings}, we conclude that no language eliminates the redundant operation.
Since this experiment is particularly simple in terms of analysis and substitution, there is no reason to explore more advanced and frequently occurring scenarios, such as the stochastic Newton method mentioned above, or the Kalman filter and signal processing shown in Table~\ref{tab:LAMP_examples}.

\begin{table}
  \centering
  \subfloat[Common Subexpression Elimination.]{
    \begin{tabular}{p{1.9cm}p{2.3cm}}
      \toprule
      Naive & Recommended \\
      \midrule
      \texttt{X := ABAB} &
      \texttt{M := AB\newline X := MM\newline} \\
      \bottomrule
    \end{tabular}
    \label{tab:CSE}
  }
  \subfloat[Loop-Invariant
            Code Motion.]{
    \begin{tabular}{p{3.2cm}p{3.2cm}}
      \toprule
      Naive & Recommended \\
      \midrule
      \texttt{for i in 1:n\newline\phantom{xx}M := AB\newline\phantom{xx}X[i] := M[i, i]} &
      \texttt{M := AB\newline for i in 1:n\newline\phantom{xx}X[i] := M[i, i]} \\
      \bottomrule
    \end{tabular}
  \label{tab:LICM}
  }
  \caption{Input expressions for Experiments \texttt{\#}9: Common Subexpression Elimination, and \texttt{\#}10: Loop-Invariant Code Motion.}
  \label{tab:CSE_LICM}
\end{table}

\subsection{Loop-Invariant Code Motion}

Another common scalar optimization is Loop-Invariant Code Motion.
For this, the compiler first looks for expressions that occur within a loop but yield the same result regardless of how many times the loop is executed~\cite[p. 592]{aho2007compilers}, and then moves them out of the loop body.
Again, when dealing with matrix computations this optimization is particularly important due to their high computational cost.
However, in the case of matrices, memory limitations might occur more frequently, compared to scalars, if intermediate storage matrices are large.
The two code snippets shown in Table~\ref{tab:LICM} extract the diagonal of the matrix product $A B$.
In the ``naive'' column, $A B$ is recomputed in every iteration of the loop, for a total cost of \order{n^{4}} floating point operations.
In the ``recommended'' column, the product is computed only once, outside the loop body, for a total cost of \order{n^{3}} operations.

\subsubsection{Experiment \texttt{\#}10: Loop-Invariant Code Motion}
\paragraph{Input}
To identify whether or not any of the languages performs loop-invariant code motion, we measured the execution time of
the snippets in Table~\ref{tab:LICM}.
\paragraph{Results}
No language eliminates the redundant operations.

\subsection{Partitioned/Blocked Operands}
In many applications such as finite element methods~\cite{blockedfiniteelements2010} and signal processing~\cite{blockedsignalprocessing2011},
matrices exhibit blocked structures (e.g.~block diagonal, block tridiagonal, block Toeplitz).
In these cases, a blocked matrix representation is often extremely convenient to write concise equations;
see Eqn.~\eqref{eq:partioned_matrices} for an example.
However, an evaluation of such expressions that does not explicitly consider the blocked structure is likely to lead to suboptimal performance.
The ability to handle each block individually can improve performance by reducing the overall amount of computation
and/or by using specialized functions on blocks with certain properties.
\begin{equation}
\begin{bmatrix}
A_1 & 0 \\
0 & A_2
\end{bmatrix}^{-1}
B =
\begin{bmatrix}
A_1 & 0 \\
0 & A_2
\end{bmatrix}^{-1}
\begin{bmatrix}
B_{T} \\
B_{B}
\end{bmatrix}
=
\begin{bmatrix}
A_1^{-1} B_{T} \\
A_2^{-1} B_{B}
\end{bmatrix}
\label{eq:partioned_matrices}
\end{equation}

\subsubsection{Experiment \texttt{\#}11: Blocked Matrices}

\paragraph{Input}
Equation~\ref{eq:partioned_matrices} shows the experiment, where a block diagonal matrix consisting of matrices $A_1,
A_2 \in $ \dims{n}{n} is used to solve a linear system.
Since all the languages considered offer mechanisms to construct matrices out of
blocks---the ``{\tt [\ ]}'' brackets in Matlab/Octave and Julia; explicit functions in Armadillo, Eigen, NumPy and R---we
examine if the structure of a blocked matrix (built with these mechanisms) is considered in subsequent operations.
To give languages the best chance, we create the blocked matrix and immediately (without temporary storage) use it to solve a linear system.
For example, the input expression for Matlab is ``\verb|C = [A1 zeros(n, n); zeros(n, n) A2]\B;|''

\paragraph{Results}
By comparing the time to solve a linear system with a blocked matrix with the time it takes to solve two small linear systems,
we conclude that no language makes use of the block diagonal structure of the matrix.

\begin{savenotes}
\begin{table}
    \centering
    \begin{tabular}{lccccccc}
    	\toprule
    	Experiment         & Armadillo & Eigen & Julia & Matlab & NumPy & Octave &  R  \\ \midrule
    	\verb|(A+B)(c, c)| &    \na    & \yes  &  \no  &  \na   &  \no  &  \no   & \no \\
    	\verb|(A+B)(:, c)| &    \na    & \yes  &  \no  &  \na   &  \no  &  \no   & \no \\
    	\verb|diag(A+B)|   &   \yes    & \yes  &  \no  &  \no   &  \no  &  \no   & \no \\ \midrule
    	\verb|(A*B)(c, c)| &    \na    &  \no  &  \no  &  \na   &  \no  &  \no   & \no \\
    	\verb|(A*B)(:, c)| &    \na    &  \no  &  \no  &  \na   &  \no  &  \no   & \no \\
    	\verb|diag(A*B)|   &    \changed{\yes}\footnote{\changed{Feature added in latest version (since 9.850-RC1). Not present in version 9.800.x.}}  & \yes  &  \no  &  \no   &  \no  &  \no   & \no \\ \bottomrule
    \end{tabular}
    \caption{Experiment \texttt{\#}12: Partial Operand Access. \changed{Armadillo and Eigen are the only languages that
        avoid unnecessary computations when a user requests the diagonal of a matrix product.}}
    \label{tab:partial_operand_access}
\end{table}
\end{savenotes}

\subsection{Partial Operand Access}
It is often the case that only parts of the output operands are needed.
Potentially, this means that not all operations need to be performed, but only those that contribute to the result.
For instance, in audio segmentation~\cite{foote2000}, the self-similarity matrix of a signal is convoluted with a kernel,
but only the elements of the diagonal are needed for further computations.

\subsubsection{Experiment \texttt{\#}12: Partial Operand Access}
\paragraph{Input}
We perform six experiments on each language to determine the extent to which this optimization is applied.
Specifically, we choose two operations, matrix addition and matrix multiplication, and request one single element, one column,
or the diagonal of the result.
The exact expressions used (for Octave) are shown in Table~\ref{tab:partial_operand_access}.

\paragraph{Results} 
As Table~\ref{tab:partial_operand_access} indicates, Eigen is the only language that fully supports this optimization for the case of matrix addition.
In the case of matrix multiplication, \changed{Armadillo and Eigen are the only languages to support the extraction of the diagonal of a product of two matrices,
without performing a matrix-matrix multiplication.}
For the other two operations, no language is able to simplify computations and avoid performing a \texttt{GEMM} before extracting the user-requested part of the result.
While Armadillo performs the optimization for the experiments \verb|diag(A+B)| \changed{ and \texttt{diag(A*B)}}, it does not support the syntax necessary for the other four experiments.
Specifically, the expressions \verb|(A+B)(c, c)| and \verb|(A+B).col(c)|, for an arbitrary constant \verb|c|, do not compile.
Similarly, Matlab does not support the indexing of the result of operations in parenthesis.

\section{Conclusions}
\label{sec:conclusions}
We consider LAMP, the problem of mapping target linear algebra expressions onto a set of available instructions while minimizing a cost function.
We provide a definition to LAMP that unifies diverse and seemingly distant research directions in numerical linear algebra and high-performance computing; from this, we prove that in general, LAMP is at
least NP-complete.
We then focus on matrix expressions that arise in practical applications, select popular programming languages that offer a high-level interface to linear algebra, and set out to investigate how efficiently they solve LAMP.
To this end, we create a benchmark consisting of simple tests, and exposing individual optimizations that are necessary to achieve good performance; these include both standard compiler optimizations such as common subexpression elimination and loop-invariant code motion, as well as linear algebra specific optimizations such as the matrix chain problem, and matrix properties.
We discuss the details of each optimization and demonstrate its effect on performance.
This investigation aims not only to showcase the capabilities and limitations of high-level languages for matrix computations, but also to serve as a guide for the future development of such languages.

\paragraph{Future Work}
The experiments included in our benchmark are meant to expose an initial set of optimizations that we deemed essential
for programming languages to generate solutions that are competitive with those created by human experts.
Other optimizations were not considered either because they do not arise so frequently,
or because in our opinion are still out of reach for modern programming languages.
For instance, with suitable assumptions of storage and dependencies,
the loop \changed{\texttt{for i in 1:n\ \ Y.col(i) = A*B.col(i)}} should be turned into \texttt{Y = A*B}, to combine multiple \changed{low-efficiency} {\tt GEMV}s into one single \changed{high-efficiency} {\tt GEMM}.

\changed{
  A natural extension to this investigation would be the solution of LAMPs in the presence of one or more forms of
  parallelism (multi-threading, accelerators, distributed-memory computations). 
  However, due to the rather limited support offered by the languages and the extra complexity of such a study, we decided to
  restrict our experiments to a sequential execution. 

  Indeed, in the case of distributed-memory computations and execution on GPUs,
  native support (not requiring the installation of third-party packages/libraries) from the target languages is lacking:
  Only MATLAB supports both types of parallelism, while Julia supports only distributed-memory computations.
  Also, in both these languages, users are required to explicitly allocate operands according to the chosen type of
  parallelism, and to explicitly express the portion of computation to be performed on each device. This manual setup 
  thus lies outside the spirit of this manuscript, i.e., an investigation of how high-level languages
  automatically solve LAMPs. 

  When it comes to multi-threading, the situation is somewhat different.
  We identified that all languages support a multi-threaded execution---at least by invoking a multi-threaded
  BLAS library---and that in fact all languages make use of multi-threading on individual BLAS kernels. 
  However,
  since it might be that different portions of a LAMP can be computed independently from one another, i.e., simultaneously,
  the optimal solution entails deciding if and how the available resources have to be split and allocated to the
  different portions.
  In order to solve this allocation problem, 
  languages would have to rely not only on the efficiency and the scalability of different
  kernels, but also on information about conflicts caused by concurrent execution of multiple kernels. 
  For all these reasons we believe that the parallel solution of LAMPs 
  merits a separate and thorough investigation on its own.

  In our experiments, we only used dense matrices. 
  Another obvious extension to this study would consider any combination of dense and sparse operands. 
  However, while all of the target languages
  support sparse matrix \emph{representation}, the assessment of how well they automatically deal with sparsity is not
  as straightforward as their dense counterpart.
  On the one hand, there exist many sparse storage formats
  which are tailored to specific applications and operations;~\cite{sparse_storage_formats}
  each of the languages we examine supports a subset of those formats.
  On the other hand, given the different features and structure of matrices from different application domains,  
  it is unclear what one should expect from a programming language if no human annotations are provided.
}

\changed{One further investigation, motivated by applications such as machine learning which require mixed precision computations,
  would repeat the experiments using
  different data types (single precision, bfloat etc.) while monitoring whether or not performance scales
  accordingly}. Finally, in all our
experiments, we concerned ourselves only with performance; in practice, numerical stability and the proper handling of
ill-conditioned matrices are critically important aspects of matrix computations.  Further experiments should be
designed to assess how high-level languages deal with such issues.


%

\section*{Acknowledgments}
Financial support from the Deutsche Forschungsgemeinschaft (German Research Foundation) through grants GSC 111 and IRTG 2379 is gratefully acknowledged.

\bibliographystyle{siamplain}
\bibliography{bibliography}

\appendix
\section{Example Problems}
\label{sec:exampleproblems}
\tcbset{ams nodisplayskip, width=0.9\textwidth, boxrule=0pt, colback=rwthverylightgray, arc=0pt, auto outer arc, left=0pt, right=0pt, boxsep=0pt}
\newcommand{\exprob}[3]{
\item #1
  \begin{tcolorbox}
    #2
  \end{tcolorbox}
  \begin{flushright}
    #3
  \end{flushright}            
}

\begin{enumerate} \setlength{\itemsep}{6pt}

\exprob{Standard Least Squares}{
\[b := (X^T X)^{-1} X^T y\]
}{
\exdims{X}{n}{m};
\exdims{y}{n}{1};
$n > m$
}

\exprob{Generalized Least Squares}{
\[b := (X^T M^{-1} X)^{-1} X^T M^{-1} y\]
}{
\exdims{M}{n}{n}, \SPD;
\exdims{X}{n}{m};
\exdims{y}{n}{1};
$n > m$
}

\exprob{Optimization \cite{straszak2015}}{
\[x := W(A^T(AWA^T)^{-1}b-c)\]
}{
\exdims{A}{m}{n};
\exdims{W}{n}{n}, \DI, \SPD;
\exdims{b}{m}{1};
\exdims{c}{n}{1};
$n > m$
}

\exprob{Optimization \cite{straszak2015}}{
\begin{align*}
  x_f &:= W A^T (AWA^T)^{-1} (b - Ax)\\
  x_o &:= W (A^T (AWA^T)^{-1} Ax - c)
\end{align*}
}{
\exdims{A}{m}{n};
\exdims{W}{n}{n}, \DI, \SPD;
\exdims{b}{m}{1};
\exdims{c}{n}{1};
$n > m$
}

\exprob{Signal Processing \cite{ding2016}}{
\[x := ( A^{-T} B^T B A^{-1} + R^T L R )^{-1} A^{-T} B^T B A^{-1} y\]
}{
\changed{\exdims{A, B}{n}{n} Band matrices (Toeplitz);}
\changed{\exdims{R}{n-1}{n}, Upper Bidiagonal;}
\exdims{L}{n-1}{n-1}, \DI;
\exdims{y}{n}{1}
}

\exprob{Triangular Matrix Inversion \cite{bientinesi2008}}{
\begin{align*}
  X_{10} &:= L_{10} L_{00}^{-1} \\
  X_{20} &:= L_{20} + L_{22}^{-1} L_{21} L_{11}^{-1} L_{10} \\
  X_{11} &:= L_{11}^{-1} \\
  X_{21} &:= - L_{22}^{-1} L_{21}
\end{align*}
}{
\exdims{L_{00}}{n}{n}, \LT;
\exdims{L_{11}}{m}{m}, \LT;
\exdims{L_{22}}{k}{k}, \LT;
\exdims{L_{10}}{m}{n};
\exdims{L_{20}}{k}{n};
\exdims{L_{21}}{k}{m}
}

\exprob{Ensemble Kalman Filter \cite{nino2016} \label{app:ekf}}{
\[X^a := X^b + ( B^{-1} + H^T R^{-1} H )^{-1} ( Y - H X^b )\]
}{
\exdims{B}{N}{N}, \SPSD;
\exdims{H}{m}{N};
\exdims{R}{m}{m}, \SPSD;
\exdims{Y}{m}{N};
\exdims{X^b}{n}{N}
}

\exprob{Ensemble Kalman Filter \cite{nino2016}}{
\[\delta X := \left( B^{-1} + H^T R^{-1} H \right)^{-1} H^T R^{-1} \left( Y - H X^b \right)\]
}{
see \ref{app:ekf}
}

\exprob{Ensemble Kalman Filter \cite{nino2016}}{
\[\delta X := X V^T \left( R + H X (H X)^T \right)^{-1} \left( Y - H X^b\right)\]
}{
\exdims{X}{m}{N};
see \ref{app:ekf}
}

\exprob{Image Restoration \cite{Tirer:2017uv} \label{app:ir}}{
\[x_k := (H^T H + \lambda \sigma^{2} I_{n} )^{-1} ( H^T y +\lambda \sigma^{2}(v_{k-1} - u_{k-1}) )\]
}{
\exdims{H}{m}{n};
\exdims{y}{m}{1};
\exdims{v_{k-1}}{n}{1};
\exdims{u_{k-1}}{n}{1};
$\lambda > 0$;
$\sigma > 0$;
$n > m$
}

\exprob{Image Restoration \cite{Tirer:2017uv}}{
\begin{align*}
  H^\dag &:= H^T ( H H^T )^{-1} \\
  y_k &:= H^\dag y + ( I_n - H^\dag H ) x_k
\end{align*}
}{
\exdims{H^\dag}{n}{m};
see \ref{app:ir}
}

\exprob{Randomized Matrix Inversion \cite{Gower:2017bq} \label{app:rmi}}{
\[X_{k+1} := X_k + W A^T S (S^T AWA^T S)^{-1} S^T (I_n - A X_k)\]
}{
\exdims{W}{n}{n}, \SPD;
\exdims{S}{n}{q};
\exdims{A}{n}{n};
\exdims{X_k}{n}{n};
$q \ll n$
}

\exprob{Randomized Matrix Inversion \cite{Gower:2017bq}}{
\begin{align*}
  \Lambda &:= S (S^T A^T W A S)^{-1} S^T \\
  X_{k+1} &:= X_k + (I_n - X_k A^T) \Lambda A^T W
\end{align*}
}{
see \ref{app:rmi}
}

\exprob{Randomized Matrix Inversion \cite{Gower:2017bq}}{
\begin{align*}
  \Lambda &:= S (S^T A W A S)^{-1} S^T \\
  \Theta &:= \Lambda AW \\
  M_k &:= X_k A - I \\
  X_{k+1} &:= X_k  - M_k \Theta - (M_k \Theta)^T + \Theta^T (A X_k A - A) \Theta
\end{align*}
}{
\exdims{A}{n}{n}, \SYM;
\exdims{X_k}{n}{n}, \SYM;
\exdims{\Lambda}{n}{n}, \SYM;
\exdims{\Theta}{n}{n};
\exdims{M_k}{n}{n};
see \ref{app:rmi}
}

\exprob{Randomized Matrix Inversion \cite{Gower:2017bq}}{
\[X_{k+1} := S(S^T A S)^{-1} S^T +  (I_n - S(S^T A S)^{-1} S^T A) X_k (I_n - A S(S^T A S)^{-1} S^T)\]
}{
\exdims{A}{n}{n}, \SPD;
\exdims{W}{n}{n}, \SPD;
\exdims{S}{n}{q};
\exdims{X_k}{n}{n};
$q \ll n$
}

\exprob{Stochastic Newton \cite{Chung:2017ws} \label{app:sn}}{
\[B_k := \frac{k}{k-1}B_{k-1} (I_n - A^T W_k ((k-1)I_l + W_k^T A B_{k-1} A^T W_k)^{-1} W_k^T A B_{k-1} )\]
}{
\exdims{W_k}{m}{l};
\exdims{A}{m}{n};
\exdims{B_k}{n}{n}, \SPD;
$l < n \ll m$
}

\exprob{Stochastic Newton \cite{Chung:2017ws}}{
\[B_1 := \frac{1}{\lambda_1} (I_n - A^T W_1 (\lambda_1 I_l + W_1^T A A^T W_1)^{-1} W_1^T A  )\]
}{
see \ref{app:sn}
}

\exprob{Tikhonov regularization \cite{Golub:2006hl} \label{app:tr}}{
\[x := (A^T A + \Gamma^T \Gamma)^{-1} A^T b\]
}{
\exdims{A}{n}{m};
\exdims{\Gamma}{m}{m};
\exdims{b}{n}{1}
}

\exprob{Tikhonov regularization \cite{Golub:2006hl}}{
\[x := (A^T A + \alpha^2 I)^{-1} A^T b\]
}{
$\alpha > 0$;
see \ref{app:tr}
}

\exprob{Generalized Tikhonov regularization \label{app:gtr}}{
\[x := (A^T P A + Q)^{-1} (A^T P b + Q x_0 )\]
}{
\exdims{P}{n}{n}, \SPSD;
\exdims{Q}{m}{m}, \SPSD;
\exdims{x_0}{m}{1};
\exdims{A}{n}{m};
\exdims{\Gamma}{m}{m};
\exdims{b}{n}{1}
}

\exprob{Generalized Tikhonov regularization}{
\[x := x_0 + (A^T P A + Q)^{-1} (A^T P ( b - A x_0 ))\]
}{
see \ref{app:gtr}
}

\exprob{LMMSE estimator \cite{Kabal:2011wr} \label{app:lmmse}}{
\[x_\text{out} = C_X A^T (A C_X A^T + C_Z)^{-1} (y - A x) + x\]
}{
\exdims{A}{m}{n};
\exdims{C_X}{n}{n}, \SPSD;
\exdims{C_Z}{m}{m}, \SPSD;
\exdims{x}{n}{1};
\exdims{y}{m}{1}
}

\exprob{LMMSE estimator \cite{Kabal:2011wr}}{
\[x_\text{out} := (A^T C_Z^{-1} A + C_X^{-1})^{-1} A^T C_Z^{-1} (y - A x) + x\]
}{
see \ref{app:lmmse}
}

\exprob{LMMSE estimator \cite{Kabal:2011wr}}{
\begin{align*}
  K_{t+1} &:= C_{t} A^T (A C_t A^T + C_z)^{-1} \\
  x_{t+1} &:= x_{t} + K_{t+1} (y - A x_{t}) \\
  C_{t+1} &:= (I - K_{t+1} A) C_{t}
\end{align*}
}{
\exdims{A}{m}{n};
\exdims{K_{t+1}}{m}{m};
\exdims{C_t}{n}{n}, \SPSD;
\exdims{C_Z}{m}{m}, \SPSD;
\exdims{x_t}{n}{1};
\exdims{y}{m}{1}
}

\exprob{Kalman Filter \cite{Kalman:1960ii}}{
\begin{align*}
K_k   &:= P_{k-1} H^{T} ( H_{k} P_{k-1} H_k^T + R_k )^{-1} \\
P_k   &:= \left( I - K_k H_k \right) P_{k-1} \\
x_{k} &:= x_{k-1} + K_k ( z_k - H_{k} x_{k-1} )
\end{align*}}{
\exdims{K_k}{n}{m};
\exdims{P_k}{n}{n}, \SPD;
\exdims{H_k}{m}{n}, \SPD;
\exdims{R_k}{m}{m}, \SPSD;
\exdims{x_k}{n}{1};
\exdims{z_k}{m}{1}
}
\end{enumerate}

\section{Timings \changed{and operand sizes}}
\label{app:timings}

\begin{table}[H]
    \centering
    \begin{adjustbox}{max width=\textwidth}
        \begin{tabular}{l@{\hspace*{2pt}}r@{\hspace*{2pt}}l@{\hspace*{2pt}}l@{\hspace*{2pt}}lcccccccc}
        	\toprule
        	\multicolumn{5}{l}{Expression}                        &  C   & Armadillo & Eigen & Julia & Matlab & NumPy & Octave &  R   \\ \midrule
        	$C_2$         &      $:=$ & $C_1$    & $+$ & $C_2$    &      &   0.01    & 0.01  & 0.03  &  0.03  & 0.03  &  0.03  & 0.03 \\ \midrule
        	$C$           &      $:=$ & $AB$     &     &          & 1.43 &   1.43    & 1.46  & 1.44  &  1.44  & 1.48  &  1.48  & 1.47 \\
        	$C$           &      $:=$ & $AB$     & $+$ & $C$      & 1.43 &   1.46    & 1.46  & 1.47  &  1.45  & 1.49  &  1.51  & 1.5  \\
        	$C$           & $\pluseq$ & $AB$     &     &          & \na  &   1.43    & 1.46  & 1.47  &  \na   & 1.49  &  1.51  & \na  \\ \midrule
        	$C$           &      $:=$ & $AA^{T}$ &     &          & 0.73 &   0.74    &       & 0.76  &  0.76  & 0.78  &  0.78  &      \\
        	$C$           &      $:=$ & $AA^{T}$ & $+$ & $C$      & 0.73 &   0.77    &       & 0.80  &  0.77  & 0.79  &  0.81  &      \\
        	$C$           & $\pluseq$ & $AA^{T}$ &     &          & \na  &   0.77    &       & 0.81  &  \na   & 0.79  &  0.81  &      \\ \bottomrule
        \end{tabular}
    \end{adjustbox}
    \caption{Experiment \texttt{\#}4: Update of C.}
    \label{tab:app_updatec}
\end{table}

\begin{table}[H]
    \begin{adjustbox}{max width=\textwidth}
        \begin{tabular}{llcccccccc}
        	\toprule
        	     \multicolumn{2}{c}{Experiment}      &       &           &       &       &        &       &        &      \\
        	\cmidrule(r){1-2} Operation & Property   &   C   & Armadillo & Eigen & Julia & Matlab & NumPy & Octave &  R   \\ \midrule
        	Multiplication              & General    & 1.46  &   1.43    & 1.46  & 1.44  &  1.44  & 1.45  &  1.48  & 1.47 \\
        	                            & Triangular & 0.73  &   1.44    & 1.46  & 1.44  &  1.44  & 1.48  &  1.48  & 1.47 \\
        	                            & Diagonal   & 0.06  &   1.43    & 1.46  & 1.44  &  1.44  & 1.48  &  1.48  & 1.47 \\ \midrule
        	Linear System               & General    & 0.65  &   0.68    &       & 0.68  &  0.74  & 0.71  &  0.77  & 0.73 \\
        	                            & Symmetric  & 0.52  &   0.68    &       & 0.68  &  0.76  & 0.71  &  0.77  & 0.73 \\
        	                            & SPD        & 0.36  &   0.42    &       & 0.65  &  0.46  & 0.68  &  0.47  & 0.70 \\
        	                            & Triangular & 0.05  &   0.09    &       & 0.06  &  0.06  & 0.71  &  0.07  & 0.70 \\
        	                            & Diagonal   & 0.002 &   0.04    &       & 0.02  &  0.06  & 0.68  &  0.08  & 0.69 \\ \bottomrule
        \end{tabular}
    \end{adjustbox}
    \caption{Experiments \texttt{\#}7--8 Properties in Multiplication and Linear Systems.}
    \label{tab:app:solve_mult}
\end{table}

\begin{table}[H]
    \centering
    \begin{adjustbox}{max width=\textwidth}
        \begin{tabular}{lcccccccc}
        	\toprule
        	Experiment  &  C   & Armadillo & Eigen & Julia & Matlab & NumPy & Octave &  R   \\ \midrule
        	naive       & 2.87 &   2.90    & 2.89  & 2.91  &  2.90  & 2.97  &  2.99  & 2.96 \\
        	recommended & 1.44 &   1.44    & 1.47  & 1.47  &  1.45  & 1.50  &  1.50  & 1.50 \\ \bottomrule
        \end{tabular}
    \end{adjustbox}
    \caption{Experiment \texttt{\#}9: Common Subexpression Elimination.}
    \label{tab:app:cse}
\end{table}

\begin{table}[H]
    \centering
    \begin{adjustbox}{max width=\textwidth}
        \begin{tabular}{lccccccc}
        	\toprule
        	Experiment  & Armadillo & Eigen & Julia & Matlab & NumPy & Octave &   R   \\ \midrule
        	naive       &   0.457   & 0.489 & 0.461 & 0.458  & 0.473 & 0.473  & 0.545 \\
        	recommended &   0.002   & 0.002 & 0.002 & 0.002  & 0.002 & 0.005  & 0.002 \\ \bottomrule
        \end{tabular}
    \end{adjustbox}
    \caption{Experiment \texttt{\#}10: Loop Invariant Code Motion.}
    \label{tab:app:licm}
\end{table}

\begin{table}[H]
    \centering
    \begin{adjustbox}{max width=\textwidth}
        \begin{tabular}{lccccccc}
        	\toprule
        	Experiment         & Armadillo & Eigen & Julia & Matlab & NumPy & Octave &  R   \\ \midrule
        	compact            &   2.06    & 2.13  & 2.08  &  2.18  & 2.23  &  2.23  & 2.17 \\
        	blocked (manually) &   0.97    & 0.98  & 0.97  &  1.01  & 1.06  &  1.06  & 1.02 \\ \bottomrule
        \end{tabular}
    \end{adjustbox}
    \caption{Experiment \texttt{\#}11: Partitioned Operands.}
    \label{tab:app:part_op}
\end{table}

\begin{table}[H]
    \centering
        \begin{adjustbox}{max width=\textwidth}
    \begin{tabular}{lccccccc}
    	\toprule
    	Experiment             & Armadillo &  Eigen   &  Julia   & Matlab &  NumPy   &  Octave  &    R     \\ \midrule
    	\verb|(A+B)(c,c)|      &    \na    & 0.000000 & 0.029364 &  \na   & 0.030491 & 0.030074 & 0.017350 \\
    	\verb|A(c,c)+B(c,c)|   &           & 0.000000 & 0.000000 &        & 0.000005 & 0.000043 & 0.000012 \\ \midrule
    	\verb|(A+B)(:,c)|      &    \na    & 0.000007 & 0.029320 &  \na   & 0.029229 & 0.030080 & 0.017368 \\
    	\verb|A(:,c)+B(:,c)|   &           & 0.000007 & 0.000022 &        & 0.000092 & 0.000063 & 0.000043 \\ \midrule
    	\verb|diag(A+B)|       &  0.0001   &  0.0001  &  0.0294  & 0.0291 &  0.0305  &  0.0298  &  0.0174  \\
    	\verb|diag(A)+diag(B)| &  0.0001   &  0.0001  &  0.0001  & 0.0001 &  0.0006  &  0.0001  &  0.0002  \\ \midrule
    	\verb|A*B|             &   1.43    &   1.46   &   1.44   &  1.44  &   1.48   &   1.48   &   1.47   \\
    	\verb|(A*B)(c,c)|      &    \na    &   1.45   &   1.44   &  \na   &   1.48   &   1.48   &   1.46   \\
    	\verb|(A*B)(:,c)|      &    \na    &   1.45   &   1.44   &  \na   &   1.48   &   1.48   &   1.46   \\
    	\verb|diag(A*B)|       &   0.033   &  0.026   &   1.44   &  1.44  &   1.48   &   1.48   &   1.46   \\ \bottomrule
    \end{tabular}
    \end{adjustbox}
    \caption{Experiment \texttt{\#}12: Partial Operand Access.}
    \label{tab:app:partial_operand_access}
\end{table}

\changed{
\subsection{Operand sizes}

For all experiments n = 3000.

Experiment \texttt{\#}1: GEMM:  \dimms{A,B,C}{n}{n}

Experiment \texttt{\#}2: SYRK:  \dimms{A,C}{n}{n}

Experiment \texttt{\#}3: SYR2K: \dimms{A,B,C}{n}{n}

Experiment \texttt{\#}4: Update of C: \dimms{A,B,C}{n}{n}

Experiment \texttt{\#}5: Explicit Inversion: \dimms{A}{n}{n}; \dimms{B, C}{n}{200}

Experiment \texttt{\#}6: Optimal Parenthesization:

\begin{itemize}
\item LtR: \dimms{M_1}{(n/5)}{n}; \dimms{M_2, M_3}{n}{n}

\item RtL: \dimms{M_1, M_2}{n}{n}; \dimms{M_3}{n}{(n/5)}

\item Mixed: \dimms{M_1,M_4}{n}{n}; \dimms{M_2}{n}{(n/5)}; \dimms{M_3}{(n/5)}{n}
\end{itemize}

Experiment \texttt{\#}7: Properties in Multiplication: \dimms{A,B,C}{n}{n}

Experiment \texttt{\#}8: Properties in Linear Systems: \dimms{A}{n}{n}; \dimms{B,C}{n}{200}

Experiment \texttt{\#}9: Common Subexpressions: \dimms{A,B,X,M}{n}{n}

Experiment \texttt{\#}10: Loop Invariant Code Motion: \dimms{A,B,M}{n}{n}

Experiment \texttt{\#}11: Blocked Matrices: \dimms{A_1,A_2}{(n/4)}{(n/4)}; \dimms{B, C}{(n/2)}{(n/2)}

Experiment \texttt{\#}12: Partial Operand Access: \dimms{A,B}{n}{n}

}

\end{document}